\documentclass[
showpacs,
floatfix,
aps,
prl,
twocolumn,
superscriptaddress,
amssymb,
]{revtex4-1}

\usepackage{amssymb}
\usepackage{latexsym}
\usepackage[dvips]{graphicx}
\usepackage{amsmath}
\usepackage{graphicx}
\usepackage{dcolumn}
\usepackage{amsfonts}
\usepackage{bm}
\usepackage{epsfig}

\newcommand{\be}{\begin{equation}}
\newcommand{\ee}{\end{equation}}
\newcommand{\bea}{\begin{eqnarray}}
\newcommand{\eea}{\end{eqnarray}}

\def\nr{n_{\mathrm{2D}}}

\def\mus{\mu_S}
\def\mul{\mu_L}

\def\rmin{\rho_{\min}}
\def\rstar{\rho_{\star}}

\newcommand{\req}[1]{Eq.\,(\ref{#1})}

\newcommand{\rfig}[1]{Fig.\,\ref{#1}}
\newcommand{\rFig}[1]{Figure \,\ref{#1}}

\newcommand{\rref}[1]{Ref.\,\onlinecite{#1}}
\newcommand{\rrefs}[2]{Refs.\,\onlinecite{#1},\,\onlinecite{#2}}

\begin{document}
\title{
Colossal negative magnetoresistance in a 2D electron gas
}
\author{Q.~Shi}
\affiliation{School of Physics and Astronomy, University of Minnesota, Minneapolis, Minnesota 55455, USA}
\author{P.~D.~Martin}
\affiliation{School of Physics and Astronomy, University of Minnesota, Minneapolis, Minnesota 55455, USA}
\author{Q.~A.~Ebner}
\affiliation{School of Physics and Astronomy, University of Minnesota, Minneapolis, Minnesota 55455, USA}
\author{M.~A.~Zudov}
\email[Corresponding author: ]{zudov@physics.umn.edu}
\affiliation{School of Physics and Astronomy, University of Minnesota, Minneapolis, Minnesota 55455, USA}
\author{L.~N.~Pfeiffer}
\affiliation{Department of Electrical Engineering, Princeton University, Princeton, New Jersey 08544, USA}
\author{K.~W.~West}
\affiliation{Department of Electrical Engineering, Princeton University, Princeton, New Jersey 08544, USA}

\received{March 5, 2014}

\begin{abstract}
We report on a colossal negative magnetoresistance (MR) in GaAs/AlGaAs quantum well which, at low temperatures, is manifested by a drop of the resistivity by more than \emph{an order of magnitude} at a magnetic field $B \approx 1$ kG.  
In contrast to MR effects discussed earlier, the MR reported here is {\em not} parabolic, even at small $B$, and persists to much higher in-plane magnetic fields and temperatures.
Remarkably, the temperature dependence of the resistivity at $B \approx 1$ kG is {\em linear} over the entire temperature range studied (from 1 to 30 K) and appears to coincide with the high-temperature limit of the zero-field resistivity, hinting on the important role of acoustic phonons.

\end{abstract}
\maketitle

%\section{Introduction}
One of the most interesting, and perhaps the most studied, properties of two-dimensional electron systems (2DES) is the magnetoresistance (MR), i.e., the change of the resistivity $\rho$ from its zero-field value $\rho_0$ due to applied perpendicular magnetic field $B$. 
At high $B$, the energy spectrum is quantized into Landau levels and MR exhibits well-known Shubnikov-de Haas oscillations and quantum Hall effects \cite{klitzing:1980,tsui:1982}. 
However, significant MR often exists even at low $B$, where quantization is not yet important.

While \emph{negative} MR ($\delta\rho = \rho(B) -\rho_0 < 0$) has been known for three decades \citep{paalanen:1983,choi:1986,li:2003}, recent studies using high mobility ($\mu \sim 10^6 - 10^7$ cm$^2$/Vs) 2DES \citep{dai:2010,dai:2011,hatke:2011b,bockhorn:2011,hatke:2012a,bockhorn:2014} presented a challenge to both quantum and quasiclassical theories. 
Although quantum theories, considering electron-electron interactions \citep{altshuler:1985,girvin:1982,gornyi:2003,gornyi:2004}, explained MR in low-mobility 2DES \cite{li:2003}, the predicted MR is way too small to explain experiments on high-mobility 2DES \citep{dai:2010,dai:2011,hatke:2011b,bockhorn:2011,hatke:2012a,bockhorn:2014}. 

Quasiclassical theories, on the other hand, can, at least in principle, produce strong negative MR in high-mobility 2DES. 
These theories consider memory effects, occurring because the probability of an electron to experience multiple collisions with the same impurity increases with $B$ and, as a result, the probability for an electron to scatter off different impurities is reduced.
The low-temperature mobility (at $B=0$) can be expressed as $\mu^{-1} = \mul^{-1} +\mus^{-1}$, where $\mul$ and $\mus$ account for scattering off \emph{long-range} (smooth) disorder, e.g. from remote ionized impurities, and \emph{short-range} (sharp) disorder, e.g. from residual background impurities, respectively.
While quasiclassical MR is the strongest in the limit of purely sharp disorder \cite{baskin:1978,bobylev:1995,dmitriev:2001,dmitriev:2002},  it can also be significant in the case of \emph{mixed} disorder with $\mul \gg \mus$ \citep{mirlin:2001,polyakov:2001}.
In this case, the theory \citep{mirlin:2001,polyakov:2001} predicts initially parabolic negative MR which crosses over to a broad minimum characterized by
\be
\rstar/\rho_0 \approx \mus/\mul\ll 1\,.
\label{rmin}
\ee

The strongest negative MR reported to date, $\rstar/\rho_0 \approx 0.02$ (at $B \approx 1$ kG), was observed in 2DES with $\mu \approx 2.2\times 10^7$ cm$^2$/Vs \citep{dai:2010,note:1}. 
While \rref{dai:2010} concluded that the MR can be explained by \req{rmin}, such a scenario appears highly unlikely as it implies $\mul > 10^9$ cm$^2$/Vs, which exceeds theoretical estimates \citep{davies:1998,hwang:2008,dmitriev:2012} by a factor of 50-100 \citep{note:77}.  
Indeed, according to \rrefs{davies:1998,note:8},
\be
\mul = 16 (e/h)(k_Fd)^3/\nr\,,
\label{mul}
\ee
where $k_F = \sqrt{2\pi n_e}$ is the Fermi wavenumber, $n_e$ is the electron density, and $\nr \simeq n_e$ is the concentration of remote impurities located at a distance $d$ from the 2D channel.
For sample A used in \rref{dai:2010} ($d = 80$ nm, $n_e = 2.9 \times 10^{11}$ cm$^{-2}$), \req{mul} yields $\mul \approx 1.7 \times 10^7 \ll 10^9$ cm$^2$/Vs \citep{note:7}.
Interestingly, \req{mul} \citep{note:8} further implies that in ultra-high mobility samples $\mul \simeq \mu$ \citep{note:77} and therefore \req{rmin} should never apply. 
As a result, existing theories predict that giant MR, with $\rstar/\rho_0 \lesssim 0.1$, can occur \emph{only} in 2DES (of typical design) with $\mu \simeq \mus \lesssim 10^6$ cm$^2$/Vs, in contradiction with experiments \citep{dai:2010,dai:2011,hatke:2011b}.

In this Rapid Communication we report on a colossal negative MR effect in a moderate-mobility ($\mu \approx 10^6$ cm$^2$/Vs) 2DES hosted in a GaAs/AlGaAs quantum well.
The hallmark of this effect is a sharp drop of $\rho(B)$ followed by a saturation at $B = B_\star \approx 1$ kG near $\rstar \equiv \rho(B_\star) \approx 0.08 \rho_0$ at $T \simeq 1$ K.
Even though the condition $\rstar/\rho_0 \simeq \mus/\mul$ appears to be satisfied in our 2DES, the effect cannot be explained by \rref{mirlin:2001}.
In particular, the low-$B$ MR correction, $-\delta\rho(B) = \rho(B)-\rho_0>0$, is found to increase roughly as $B^{1.4}$, in contrast to $B^2$ found in both theory \citep{dmitriev:2001,mirlin:2001,cheianov:2004} and recent experiments \citep{dai:2010,dai:2011,hatke:2012a,bockhorn:2011,bockhorn:2014}.
Furthermore, the MR in our 2DES remains essentially unaffected by very strong in-plane magnetic fields, up to $B_\parallel \approx 30$ kG.
This finding contrasts with recent studies \citep{dai:2011,hatke:2012a,bockhorn:2014}, in which MR was greatly suppressed by $B_\parallel \lesssim 10$ kG. 
Finally, the MR in our 2DES remains significant up to $T =  30$ K, in contrast to \rref{hatke:2012a}, where it disappeared above $2.5$ K.
The most striking feature is that $\rstar$ increases {\em linearly} over the {\em entire} $T$ range, following $\rstar (T)= \rstar^0(1 + T/T_0)$, with $\rstar^0 \approx 1.0$ $\Omega$ and $T_0 \approx 1.1$ K.
Interestingly, $\rstar(T)$ mimics the high-$T$ limit of $\rho_0(T)$, which is known to originate from electron-phonon scattering.
Taken together, these observations suggest that we have observed a colossal negative MR effect which is distinct from the effects reported previously.

Our sample is a Hall bar of width $w=200$ $\mu$m fabricated from a symmetrically doped, 29 nm-wide GaAs/AlGaAs quantum well, with the Si $\delta$-doping layers separated from the active channel by spacers of width $d\approx 80$ nm.  
At $T \simeq 1$ K, the electron density and the mobility were $n_e \approx 2.8 \times10^{11}$ cm$^{-2}$ and 
and $\mu\approx 1.0 \times10^{6}$ cm$^2$/Vs.
The magnetoresistivity $\rho(B)$ was measured in sweeping magnetic fields by a standard four-terminal lock-in technique at temperatures up to 30 K. 

%%%%%%%%%%%%%%%%%%%%%%%%%%%%%%%%%%%

\begin{figure}[t]
\includegraphics{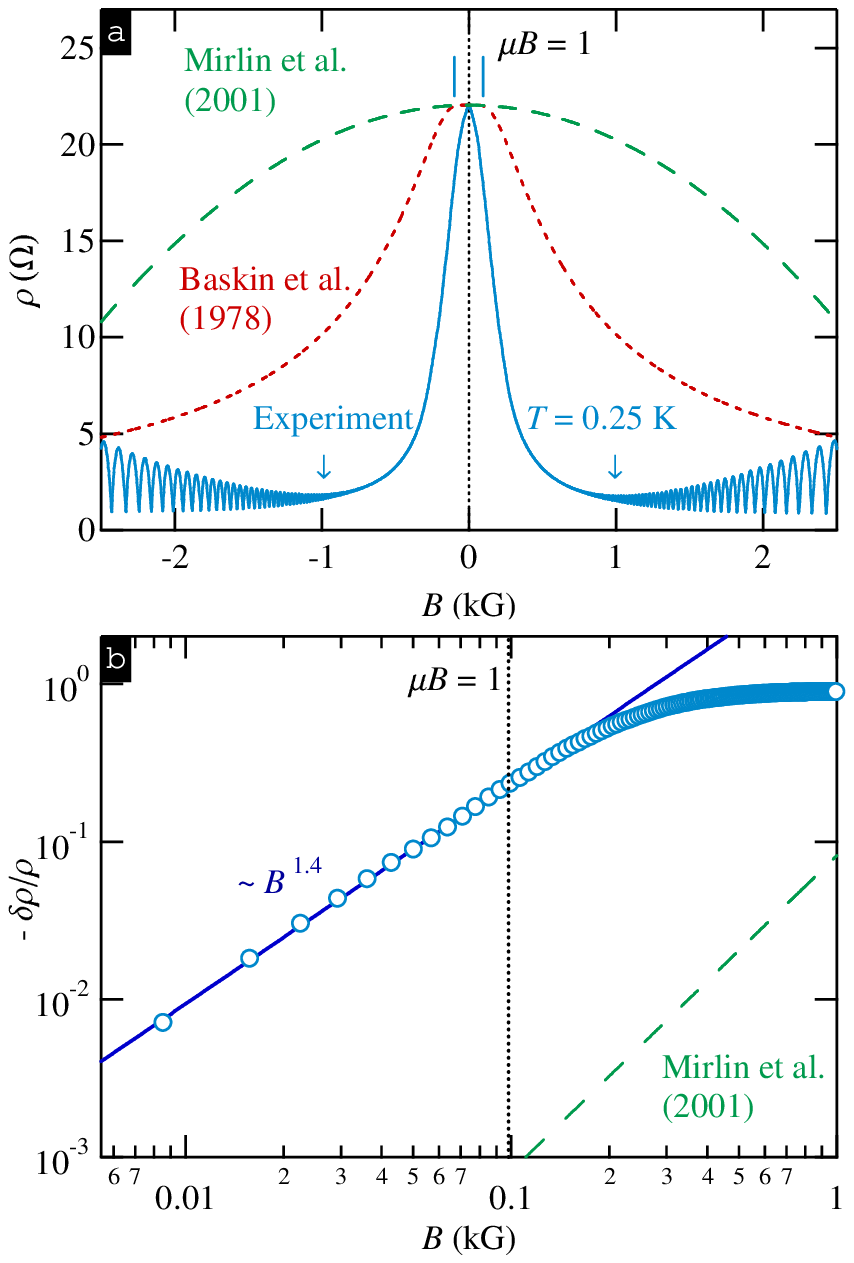}
\vspace{-0.05 in}
\caption{(Color online) 
(a) $\rho(B)$ measured at $T \simeq $ 0.25 K (solid curve), calculated according to \rref{mirlin:2001} using  $n_i$=(0.8$\mu$m)$^{-2}$, $\mus/\mul=0.1$ (dashed curve) and \rref{baskin:1978} using $\mu = \mus$ (dotted curve). 
(b) $-\delta \rho/\rho$ versus magnetic field $B$, plotted on a log-log scale. 
The fit (solid line) to the data at $B \leq$ 0.15 kG gives $-\delta \rho/\rho = (B/\bar B)^{1.4}$ with $\bar B \approx 0.28$ kG. 
For comparison, $B^2$-dependence according to \rref{mirlin:2001} is shown by a dashed line.
}
\vspace{-0.15 in}
\label{compare}
\end{figure}

%%%%%%%%%%%%%%%%%%%%%%%%%%%%%%%%%%

In \rfig{compare}(a) we present $\rho(B)$ measured at $T$ = 0.25 K (solid curve) showing a dramatic decrease which terminates at $B = B_\star \approx$ 1 kG with $\rstar \approx 0.08 \rho_0$.
Since this value is close to $\mus/\mul \approx \mu/\mul \approx 0.07$, where $\mul \approx 1.7 \times 10^7$ cm$^2$/Vs was obtained from \req{mul}, it appears possible that our data can be explained by \rref{mirlin:2001}, which proposed \req{rmin}.
However, as we show next, the MR in our 2DES is much stronger than all existing theoretical predictions.

At low magnetic fields, for $\mu \approx \mus \ll \mul$, the theory with mixed disorder model \cite{mirlin:2001} predicts
\be 
\rho(B)/\rho_0 = 1-B^2/B_0^2\,, 
\label{mirlin}
\ee
where $B_0=(h/e) \sqrt{n_i n_e} (2 \mus/\mul)^{1/4}$ and $n_i$ is the 2D density of strong scatterers.
From $2 l a_B n_i =1$, where $l$ is the mean free path and $a_B \approx 10$ nm is the Bohr radius in GaAs, we estimate $n_i = (0.42\mu$m$)^{-2}$  and then obtain $B_0 = 3.5$ kG.
Using this value and \req{mirlin}, we calculate $\rho(B)$ and present the result in \rfig{compare}(a) (dashed line).
It is clear that there a big discrepancy exists between the theoretical and experimental $\rho(B)$.
Although the theory does predict a significant drop of $\rho(B)$, the experiment shows a much steeper drop, i.e. the MR effect develops at \emph{much} lower magnetic fields .

The negative MR in our data is even stronger than the limit of purely sharp disorder (Lorentz gas model) \cite{baskin:1978,bobylev:1995,dmitriev:2001}, $\mu=\mus$, which predicts the largest possible negative MR due to classical memory effects.   
According to this model, $\rho(B)/\rho_0$ is given by
\be
\rho(B)/\rho_0 = 1-e^{-2\pi/\mu B}\,,
\label{baskin}
\ee
where $e^{-2\pi/\mu B}$ is the fraction of electrons which complete cyclotron orbits without colliding with impurities and thus do not contribute to the resistivity. 
While this simple model underestimates the MR at very low $B$ \cite{dmitriev:2001}, the difference between \req{baskin} and our data remains significant even at high $B$, see \rfig{compare}(a). 
Indeed, at $B = B_\star \approx 1$ kG, \req{baskin} gives $\rho/\rho_0 \approx 0.5$, almost an order of magnitude higher than our data.

We further demonstrate that MR in our 2DES is not quadratic in $B$, in contrast to both present theoretical \citep{dmitriev:2001,mirlin:2001,cheianov:2004} and recent experimental \citep{dai:2010,dai:2011,hatke:2012a,bockhorn:2011,bockhorn:2014,note:19} studies.
In \rfig{compare}(b) we plot $-\delta\rho/\rho_0$ versus $B$ on a log-log scale. 
The fit (solid line) to the lower $B$ part of the data, $B \leq$ 0.15 kG, gives $-\delta \rho/\rho_0 = (B/\bar B)^{1.4}$, with $\bar B$ = 0.28 kG. 
At higher $B$, $-\delta \rho(B)$ slows down and eventually saturates.
Comparison to the theoretical curve, \req{mirlin} (dashed line), reveals a two orders of magnitude difference at $B \approx 0.1$ kG.

%%%%%%%%%%%%%%%%%%%%%%%%%%%%%%%
\begin{figure}[t]
\includegraphics{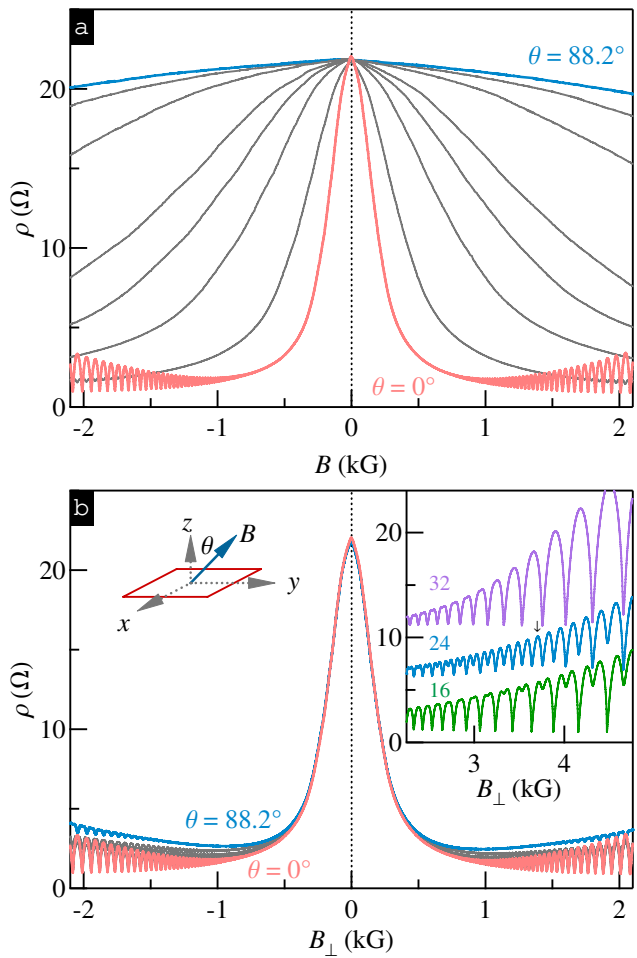}
\vspace{-0.05 in}
\caption{(Color online) 
(a) $\rho (B)$ measured at $T \simeq$ 0.25 K at different tilt angles $\theta$ = 0$^{\circ}$, 60$^{\circ}$, 75.6$^{\circ}$, 80.3$^{\circ}$, 82.8$^{\circ}$, 86.4$^{\circ}$, 87.6$^{\circ}$, 88.2$^{\circ}$.
(b) $\rho$ versus perpendicular magnetic filed $B_\perp$ at the same tilt angles.
Inset shows Shubnikov-de Haas oscillations at different values of $B/B_\perp$ which are marked by integers.
The traces are vertically offset for clarity by 4 $\Omega$.
}
\vspace{-0.15 in}
\label{tilt}
\end{figure}
%%%%%%%%%%%%%%%%%%%%%%%%%%%%%%

We next discuss the effect of an in-plane magnetic field on the colossal negative MR.
The measurements were performed with the sample tilted by angle $\theta$ with respect to the magnetic field $B$.
To facilitate the discussion, we introduce $B_\perp = B\cos \theta$ and $B_\parallel = B \sin\theta$, which denote out-of-plane and in-plane magnetic field, respectively.
\rFig{tilt}(a) shows $\rho(B)$ measured at different $\theta$ from 0$^{\circ}$ to 88.2$^{\circ}$ at $T \simeq$ 0.25 K. 
As one can see, the MR correction $\delta\rho(B)$ gets considerably smaller with increasing $\theta$, as one would expect if the MR effect is caused primarily, if not solely, by $B_\perp$.
To see if this is the case,  we present in \rfig{tilt}(b) the same data as a function of $B_\perp$. 
The inset illustrates the evolution of Shubnikov-de Haas oscillations as a result of enhanced spin splitting and effective mass renormalization in our finite-width 2DES \citep{hatke:2012c}.
Remarkably, all of the curves collapse into one universal curve demonstrating that the colossal negative MR remains essentially unchanged up to the highest angle, $\theta$ = 88.2$^{\circ}$, corresponding to $B_{\parallel}/B_\perp \approx 32$.
Indeed, even at $B_\parallel \approx 32$ kG, the drop of the resistivity is still about one order of magnitude, $\rstar/\rho_0 \approx 0.11$.
This result is vastly different from previous studies of negative MR \cite{hatke:2012a,dai:2011,bockhorn:2014}, where the negative MR was found to be strongly suppressed by $B_\parallel$ smaller than 10 kG. 

%%%%%%%%%%%%%%%%%%%%%%%%%%%%%

\begin{figure}[t]
\includegraphics{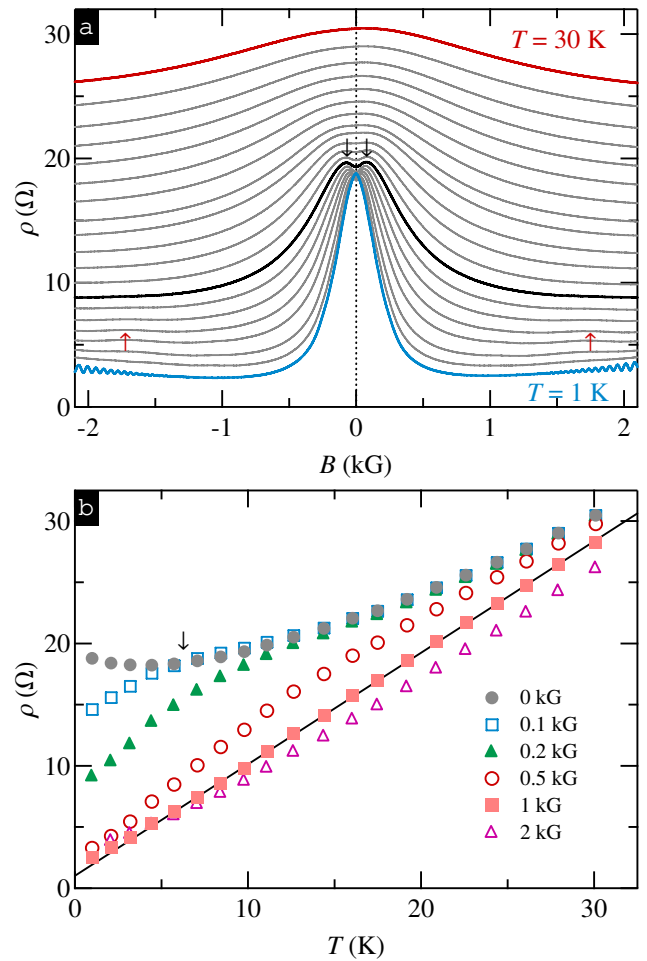}
\vspace{-0.05 in}
\caption{(Color online) 
(a) $\rho (B)$ at different temperatures $T$ from 1 to 30 K. 
Weak phonon-induced resistance oscillations can be seen around $T \approx$ 5 K (solid line), as marked by arrows. 
(b) Resistivity $\rho$ versus temperature $T$ at different magnetic fields, as marked.
 $\rho (T)$ at $B$ = 1 kG is fit by (solid line) $\rstar(T) = \rstar^0 (1+T/T_0)$, with $\rstar^0$ = 1 $\Omega$ and $T_0$ = 1.1 K.
}
\vspace{-0.15 in}
\label{tmp}
\end{figure}

%%%%%%%%%%%%%%%%%%%%%%%%%%%%%%%%%%%%55

We next discuss perhaps the most intriguing experimental aspect of this colossal negative MR effect, its temperature dependence.
In \rfig{tmp}(a) we present $\rho (B)$ at different temperatures from 1 to 30 K. 
As the temperature is elevated, MR becomes weaker but remains significant up to 30 K, in contrast to previous study \cite{hatke:2012a}, where MR virtually disappeared above 2.5 K.
We also observe signatures of phonon-induced resistance oscillations (cf. $\uparrow$), which are commonly seen in this temperature range \citep{zudov:2001b,zhang:2008,hatke:2009b,hatke:2011d,raichev:2009,dmitriev:2010} in high-mobility 2DES.
\rFig{tmp}(a) further shows that at any finite $B$, $\rho(T)$, at least initially, increases faster than $\rho_0(T)$.
At very low $B$, this behavior contributes to the development of a local minimum at $B=0$, surrounded by maxima (cf. $\downarrow$), which can be seen only at intermediate temperatures.

To examine the temperature dependence in more detail, we construct \rfig{tmp}(b) which shows $\rho$ as a function of $T$ for different $B$ from 0 to 2 kG, as marked.
Most remarkably, we find that at $B=B_\star \approx 1$ kG (solid squares), $\rho$ is a linear function of temperature over the entire range studied.
This dependence is well described by $\rstar(T) = \rstar^0 (1+T/T_0)$, with $\rstar^0$ = 1 $\Omega$ and $T_0$ = 1.1 K (solid line).    
One can notice that, initially, the resistivity at $B<B_\star$, e.g. $B = 0.5$ kG (open circles), increases at a faster rate than $\rstar(T)$, whereas at $B > B_\star$, e.g. $B = 2$ kG (open triangles), it increases at a somewhat lower rate.
At higher $T$, all of the data, including $\rho_0(T)$, converge to one common linear dependence. 
It is rather remarkable that $\rstar(T)$ is very well described by this universal dependence down to very low temperatures.
This is in vast contrast to \rref{hatke:2012a}, where at $T \lesssim 2.5$ K the $T$-dependence is superlinear for \emph{all} $B$, while at $T \gtrsim 2.5$ K there is no MR effect and all curves coincide with $\rho_0(T)$.

It is well known that the $T$-dependence of the zero-field resistivity $\rho_0$ is split into two regimes by the Bloch-Gr$\ddot{\rm u}$neisen temperature $T_{\rm BG}=2 \hbar k_F s/k_B$, where $s$ is the sound velocity.
At $T \gg T_{\rm BG}$, all phonon modes that electrons can scatter off are highly populated and $\rho(T)$ is linear, reflecting classical distribution of phonons.
At $T \ll T_{\rm BG}$, only phonons with momenta much smaller than $k_F$ are populated and $\rho(T)$ exhibits a high power-law dependence \cite{stormer:1990}.
While $\rho_0(T)$ roughly follows the expected behavior, the apparent extension of the linear dependence to low temperatures at $B = B_\star$ is totally unexpected. 
The questions one may ask are (i) how the electron-phonon scattering is modified by finite magnetic field, and (ii) how such modification translates to a change in resistivity.

It is well known that a combination of phonon-assisted electron backscattering and Landau quantization modifies electron-phonon scattering rate in a non-trivial way, leading to the $1/B$-periodic oscillations in the resistivity, occurring at $T\simeq T_{\rm BG}$ \citep{zudov:2001b,hatke:2009b,hatke:2011d,raichev:2009,dmitriev:2010}.
At the same time, it is understood that the magnetic field cannot induce any non-oscillatory correction to the electron-phonon scattering rate. 
However, different $T$-dependencies of $\rho$ observed at different $B$ do not necessarily call for different electron-phonon scattering rates.
Indeed, non-Markovian transport \cite{mirlin:2001, polyakov:2001} implies that at finite $B$ the interplay of sharp and smooth disorder is non-trivial and $\rho$ is no longer proportional to a simple sum of the corresponding scattering rates. 
Similarly, one should not expect that at finite $B$ the total scattering rate is a simple sum of rates due to disorder and phonons. 
Future theories should perhaps consider if the low-energy phonons can act similarly to a smooth disorder, i.e. effectively assisting in delocalizing electrons, which might lead to a much stronger $T$-dependence of $\rho$ at finite $B$ compared to $\rho_0(T)$. 
In fact, such a scenario has been examined in \rref{laikhtman:1994c} in the context of a smooth-disorder localization model, which predicted $\rho(B\neq 0) \propto T^\alpha$, where $\alpha$, determined by percolation scaling exponents, is lower than a power governing the $T$-dependence of $\rho_0$. 

While linear $T$-dependence of $\rho_\star$, coinciding with the high-$T$ limit of $\rho_0(T)$, strongly hints on phonons, electron-electron interactions might also be considered. 
Indeed, at low $T$, the electron-electron scattering time $\tau_{ee}$ is shorter than the electron-phonon scattering time $\tau_{ph}$.
Therefore, it should be $\tau_{ee}$, rather than $\tau_{ph}$, acting as a cutoff time, if the electrons rely on such scattering processes to transfer between different trajectories, e.g., when close to a percolation threshold.

Finally, we mention that most, if not all, observations of strong negative MR were limited to Hall bar samples \citep{dai:2010,dai:2011,hatke:2011b,bockhorn:2011,hatke:2012a,mani:2013,bockhorn:2014}.
Moreover, it was recently reported  \citep{mani:2013} that both the strength and the characteristic magnetic field of negative MR depend on the width of the Hall bar, which might indicate the importance of edge scattering or current distribution within the device.

In summary,  we have observed a colossal negative MR effect in a moderate-mobility 2DES in a GaAs/AlGaAs quantum well.
The effect is marked by a steep drop of $\rho(B)$ followed by a saturation at $B = B_\star \approx 1$ kG near $\rstar \approx 0.08 \rho_0$ at $T \lesssim 1$ K.
While the condition $\rstar/\rho_0 \simeq \mus/\mul$ seems to be satisfied in our 2DES, neither the magnitude nor the dependence $\rho(B)$ can be explained by existing theories.
More specifically, $-\delta\rho(B) = \rho(B)-\rho_0$ is found to increase as $B^{1.4}$, in contrast to results of previous theoretical \citep{dmitriev:2001,mirlin:2001,cheianov:2004} and experimental \citep{dai:2010,dai:2011,hatke:2012a,bockhorn:2011,bockhorn:2014} studies.
Furthermore, unlike previous studies \citep{dai:2011,hatke:2012a,bockhorn:2014}, the colossal MR reported here remains essentially unaffected by strong in-plane magnetic fields, up to $B_\parallel \approx 30$ kG. 
Finally, the MR in our 2DES persists up to $T =  30$ K, in contrast to \rref{hatke:2012a} where it virtually vanished at $2.5$ K.
The most remarkable feature of the observed $T$-dependence is that $\rstar(T)$ increases {\em linearly} over the {\em entire} $T$ range.
This linear dependence appears to be nearly the same as the high-$T$ limit of $\rho_0(T)$, which is well understood in terms of electron-phonon scattering.
Taken together, our findings indicate that the observed colossal negative MR is qualitatively different from the effects observed in all of the previous studies.
To identify the origin of this remarkable phenomenon further investigations are necessary.
In particular, it would be interesting to perform microwave photoresistance \citep{zudov:2001a} and nonlinear transport \citep{yang:2002} measurements which should help to better understand the correlation properties of the disorder potential \citep{dmitriev:2012} in our 2DES.

We thank S. Das Sarma, M. Dyakonov, I. Gornyi, R. Haug, M. Khodas, D. Polyakov, B. Shklovskii, and V. Umansky for discussions and G. Jones, T. Murphy, and D. Smirnov for technical assistance with experiments.
A portion of this work was performed at the National High Magnetic Field Laboratory, which is supported by NSF Cooperative Agreement No. DMR-0654118, by the State of Florida, and by the DOE and at the Center for Integrated Nanotechnologies, a U.S. Department of Energy, Office of Basic Energy Sciences user facility.   
The work at Minnesota was supported by NSF Grant No. DMR-1309578 (measurements in perpendicular fields in Minnesota) and by DOE Grant No. DE-SC002567 (tilt-field measurements at NHMFL). 
The work at Princeton was partially funded by the Gordon and Betty Moore Foundation and the NSF MRSEC Program through the Princeton Center for Complex Materials (Grant No. DMR-0819860).

%\bibliography{../../../bibRMP1,footnotes_nmr}
%\bibliography{bibRMP1,footnotes_nmr}

\begin{thebibliography}{45}
\bibitem[{not({\natexlab{c}})}]{note:8}
\bibinfo{note}{We note that \req{mul} does not take into account possible
  inter-donor correlations which might increase $\mul$.}
\end{thebibliography}

\end{document}